\documentclass[aps,prc,showpacs,nofootinbib,twocolumn]{revtex4}

\usepackage{epsfig}
\usepackage{graphicx,amsmath}
\usepackage{color}
\newcommand\ba{\begin{eqnarray}}
\newcommand\ea{\end{eqnarray}}

\newcommand{\be}{\begin{equation}}
\newcommand{\ee}{\end{equation}}

\newcommand{\bas}{\begin{eqnarray*}}
\newcommand{\eas}{\end{eqnarray*}}
\begin{document}
\title{\bf \large Relations between elementary particle masses}

\author{B. Tatischeff}
\thanks{e-mail : tati@ipno.in2p3.fr}
\affiliation{\it CNRS/IN2P3, Institut de Physique Nucl\'eaire, UMR 8608, 91405 Orsay, France, Univ. Paris-Sud, Orsay, F-91405, France}

\author{I. Brissaud}
\thanks{e-mail : brissaud.ivan@orange.fr}
\affiliation{CNRS, Honorary Research Director, France}


\vspace*{1cm}
\begin{abstract}
Relations between elementary particles masses are given using only known physical constants, without any arbitrary number.
\end{abstract}
\maketitle

The large and increasing number of hadronic particles, suggests to  find new classifications, in addition to those already existing based on their quantum numbers (isospin, spin, charge conjugation and parities). A possible approach is to look to eventual fractal properties of these particles. We recall that Nottale \cite{nottale1} noted that the "{\it lepton 
{\it e, $\mu$, $\tau$} mass ratios followed a power-law sequence, namely, 
$m_{\mu}\approx 3\ast 4.1^{3} \ast m_{e}$ = 105.656~MeV, and 
$m_{\tau}\approx 3\ast 4.1^{5} \ast m_{e}$ = 1776.1~MeV}".

In the same mind, we give relations between elementary particle masses: quarks, bosons, and (an other relation) between leptons. All experimental masses, except those specifically indicated, are taken from Review of Particle Physics (PDG) \cite{pdg}. 
\subsection{Quark masses}
 We present here power law relations between quark masses.

Six quarks are known and classified into two families. The charge of the first family
is 2/3, and the quarks are:  "u", "c", and "t". The charge of the second family is -1/3, and the quarks are:  "d", "s", and "b". These masses are:\\
m$_{u}$ = 1.5 - 3.3~MeV, we take  m$_{u}$ = 2.28~MeV\\
m$_{c}$ = 1.27 $^{+0.07}$\hspace*{-8.mm}$_{-0.11}$~GeV\\
m$_{t}$ = 171.2 $\pm$2.1GeV \cite{aaltonen1}\\

\hspace*{-4.mm}m$_{d}$ = 3.5 - 6.0~MeV, we take  m$_{d}$ = 5.1~MeV\\
m$_{s}$ = 104 $^{+ 26}$\hspace*{-5.5mm}$_{- 34}$~MeV\\
m$_{b}$ = 4.20 $^{+ 0.17}$\hspace*{-7.5mm}$_{- 0.07}$~GeV\\
  
The ratio between m$_{u}$ and m$_{d}$, r=0.45, correponds to the value used in theoretical calculations.\\
In each family, we attribute successively, the numbers 1, 2, and 3 to the three quarks.

In order to get the relations between all quark mass ratios, the remarquable ratio r  = 1/134.8 between the charm quark and the top quark masses,  suggests to use the fine structure constant $\alpha$ = e$^{2}/4 . \pi . \epsilon_{0} \hbar c$ = 1/137.036  (see \cite{aaltonen2}).
\begin{equation}
m^{(1)} _{n+1} = 2^{2(2-n)}m^ {(1)}_{n}/\alpha, 
\end{equation}
\vspace*{-6.mm}

\begin{equation}
ln\hspace*{1.mm}(m^ {(1)}_{n+1}/m^ {(1)}_{n}) = ln\hspace*{1.mm}(2^{2(2-n)}) - ln\hspace*{1.mm} \alpha.
\end{equation}

The application of the formula gives: \\
$m_{c}$ = 1.24977~GeV, and \\
$m_{t}=m_{c}/\alpha$ = 171.295~GeV.

The relation between quark masses of the second family is:
\begin{equation}
m^ {(2)}_{n+1} = (m_{\pi})/m_{p})\hspace*{1.mm}n\hspace*{1.mm} m^{(2)}_{n}/\alpha, 
\end{equation}
\vspace*{-6.mm}

\begin{equation}
ln\hspace*{1.mm}(m^{(2)}_{n+1}/m^{(2)}_{(n)}) = ln\hspace*{1.mm} ((m_{\pi}/m_{p}) n) - ln\hspace*{1.mm}\alpha
\end{equation}

where $m_{\pi}/m_{p}$ = 0.14875236 is the ratio of the pion mass to the proton mass. The application of the formula gives: $m_{s}$ = 103.961~MeV, and $m_{b}$ = 4240.0~MeV.
The masses obtained by these formulas are shown in table~I.
\begin{table}[h]
\begin{center}
\caption{Comparison of quark calculated and experimental masses (in GeV).}
\label{Table I}
\begin{tabular}[t]{c c c c}
\hline
quark&calculated&experimental&relative\\
           &mass       &mass             &difference\\
\hline
u & &0.00228&\\
c &1.25&1.27$^{+0.07}$\hspace*{-7.mm}$_{-0.11}$&1.6 10$^{-2}$\\
t &171.3 &171.2 $\pm$2.1&5.6 10$^{-4}$\\
d & &0.0051&\\
s &0.10396&0.104 $^{+0.026}$\hspace*{-7.8mm}$_{-0.034}$&3.8 10$^{-4}$\\
b &4.24&4.20 $^{+ 0.17}$\hspace*{-7.mm}$_{- 0.07}$&9.4 10$^{-3}$\\
\hline
\end{tabular}
\end{center}
\end{table}

The masses of eventual higher mass quarks, are successively for both families: 
M$^ {(1)}$ = 5865~GeV, and M$^ {(2)}$ = 257~GeV. The present experimental limit is m$_{t'}\ge$~256~GeV \cite{acosta} and m$_{b'}\ge$~128~GeV in case of (SM4) (a possible fourth generation particles \cite{holdom}).

The previous relations (1) and (3) allow to get the following relation between the masses of the first m$^{(1)}$ and the second m$^{(2)}$ quark families:
\begin{equation}
m^{(2)}_{(n+1)} / m^{(2)}_{(n)} = (m^{(1)}_{(n+1)} / m^{(1)}_{(n)})  2^{2  (n-2)} (m_{\pi} / m_{p})  n
\end{equation}
\subsection{Gauge boson masses}
These bosons are:\\
 - the photon with a mass m$_{\gamma}\le$1\hspace*{1.mm}. 10$^{-18}$~ev,\\
 - the gluon without mass,\\
 - the W, m$_{W}$ = 80.398 $\pm$ 0.025~GeV, and\\
  - the Z, m$_{Z}$ = 91.1876 $\pm$ 0.0021~GeV.

The following relation allows to get the experimental masses, with a shift from the experimental values, however much larger than the experimental precisions.
\begin{equation}
{M = m_{p} \alpha^{-1} (n/10)^{1/2}}
\end{equation}

\hspace*{-4.mm}n = 0 gives M = 0, corresponding to the photon and the gluon mass,\\
n = 4 gives M = 81.319~GeV; the relative shift from the  experimental  W mass equals 1.15 10$^{-2}$,\\
n = 5 gives M=90.917~MeV;  the relative shift from the  experimental  Z mass equals  3 10$^{-3}$ \cite{aboson}.
The masses obtained by this formula are shown in table~II.
\begin{table}[h]
\begin{center}
\caption{Comparison of gauge boson calculated and experimental masses (in GeV).}
\label{Table II}
\begin{tabular}[t]{c c c c}
\hline
gauge&calculated&experimental&relative\\
boson &mass       &mass             &difference\\
\hline
$\gamma$&0&0&\\
gluon&0&0&\\
W &81.319 &80.398 $\pm$ 0.025&1.15 10$^{-2}$\\
Z &90.917&91.1876 $\pm$ 0.0021&3 10$^{-3}$\\
\hline
\end{tabular}
\end{center}
\end{table}
It is difficult to predict a rule for the mass incrementation. Indeed we use 
"n" = 0, or 4, or 5. If we choose "n" = 9 for the next mass, {\it which can be the mass of the Higgs boson, we get M$_{H}\approx$122~GeV}. 
If we choose simply M = m$_{p} \alpha^{-1}$, we get  M$_{H}\approx$128.6~GeV. It is also possible to propose two Higgs boson masses, using the incrementation "n" = 0, 4, 5, 9, and 10, and the masses of both bosons will be:
M$_{H}\approx$~122~GeV et M$_{H}\approx$~128.6~GeV.

The Higgs boson mass limits was recently suggested to lie between 110 and 200 GeV \cite{aaltonen3} and a
  recent analysis  concluded that the Higgs boson mass should range between 115 and 148~GeV \cite{jens1}. 

We are aware of the incrementation jump, leaving out "n" = 1, (40.716~GeV), "n" = 2 (57.58~GeV), "n" = 3 (70.52~GeV), "n" = 6 (99.73~GeV), "n" = 7 (107.72~GeV), and "n" = 8 (115.16~GeV). {\it Do some of these masses correspond to still unobserved particles, in case of (SM4); and if it is the case, why do have they never been observed ? \cite{deusol}}.
\subsection{Relation between gauge boson and quark masses}
The relations between all quark masses in one hand and between all gauge boson masses in the other hand, were given previously. Therefore, we show only the relation between one gauge boson and the quark masses, more precisely  between the Z boson mass and the "u" and "c" quark masses:
\begin{equation}
m_{Z} = m_{p}/(4 \sqrt(2)) (m_{c} / m_{u}) 
\end{equation}\\
We point out that by inverting the previous relations, we get the proton mass by using the gauge boson mass(es), just as the ratio of quark masses by means of the proton and the Z boson masses, and the pion mass by means of the proton and the ratio between two quark masses. All these masses are related between them by the fine structure constant $\alpha$.\\

\subsection{Lepton masses}
A power-law-type sequence between {\it e, $\mu$}, and {\it $\tau$} leptons was given by Nottale \cite{nottale1} as already noted.

We propose here another relation, which gives less exact leptonic masses
than Nottale's relation, but which also calculates masses for two neutrinos, and 
is obtained without introduction of "external" numbers. The six leptons:
$\nu_{e}$, $\nu_{\mu}$, $\nu_{\tau}$, e, $\mu$, and $\tau$, are associated with index "k" = 1, 2, 3, 4, 5, and 6. Writing "n"= "k - 3", we have:
\begin{equation}
m^{(k)} = m_{\pi}\hspace*{1.mm}\alpha ^{(2-n)} r ^{(n(n+1)+1/2)} (2+(-1)^{n}) / h(n)
\end{equation}\\
where
\begin{equation}
 h(n) = [n^{2}- ((n+1)/4) ((-1)^{n} - 1)]\hspace*{1.cm}
\end{equation}\\
\hspace*{0.mm}where r = 1.001378 is the ratio of the neutron mass to the proton mass. It corresponds to a correction term, very close to 1 .  The h(n) term vanishes for "n = 0", preventing the determination of the  $\nu_{\tau}$ mass through such a mass relation.

The masses obtained by this formula are shown in table~III.
\begin{table}[h]
\begin{center}
\caption{Comparison of lepton calculated and experimental masses.}
\label{Table III}
\begin{tabular}[t]{c c c c}
\hline
lepton&calculated&experimental&relative\\
           &mass       &mass             &difference\\
\hline
$\nu_{e}$    &0.3~eV&0.05$\le m_{\nu_{e}}\le$0.23~eV \cite{joaquim}&\\
$\nu_{\mu}$&54~eV&$\le$~0.19~MeV&\\
$\nu_{\tau}$&        &$\le$~18.2~MeV&\\
$m_{e}$&0.511004~MeV&0.5109989~MeV~&8 .10$^{-6}$\\
$m_{\mu}$&105.62~MeV&105.65837~MeV&3.6 10$^{-4}$\\
$m_{\tau}$&1768.97~MeV&1776.84~MeV&4.4 . 10$^{-3}$\\
\hline
\end{tabular}
\end{center}
\end{table}

The calculated mass of an eventual heavier lepton is: m~$\approx$~505.5~GeV, very close to the value m~$\approx$~502~GeV found by Nottale \cite{nottale1}. The experimental lower mass limit of this heavier lepton is very unprecise, of the order of m$\ge$ 100~GeV. 
Several calculations propose an eventual fermion additional generation (see for example \cite{jens}). 

 Since the lepton masses depend only on the pion mass, and to a small amount, on the ratio of neutron over proton masses, a relation between lepton masses and quark (boson) masses is meaningless. However these masses are related to the other masses, discussed previously, through the pion mass. The relations are rather simple.\\

{\it In conclusion, we are able to rely nearly all elementary particle masses with each other, by using proton, pion, and neutron masses, and fine structure constant. These relations should be helpful for a better knowledge of particle properties. They allow to predict the masses of still unobserved elementary particles, specially possible masses for the Higgs bosons at $m_{H}$ = 122 and 128.6~GeV}.


\begin{thebibliography}{99}
\bibitem{aaltonen2}With the new top quark mass = 174.8~ Gev,  the ratio between charm and top quark masses r = 1/137.6 is still
nearer to the $\alpha$ value.
\bibitem{aboson}It is possible to reduce these shifts, by using the following two factors:\\
\hspace*{3mm} v = $(m_{\pi^{0}} /  m_{\pi^{+}})^{1/2}$ = 0.9834. A shift of  429~MeV remains therefore, between the experimental and the calculated W masses, which must be compared to the experimental imprecision: 25~MeV; and\\
\hspace*{3mm} (2 r - 1) = 1.002757.  A shift of  20~MeV remains therefore, between the experimental and the calculated Z masses, which must be compared to the experimental imprecision: 2.1~MeV.
\end{thebibliography}
\end{document}